\documentclass{aa}  
\usepackage{graphicx}
\usepackage{amsmath}
\usepackage{amssymb}
\usepackage{txfonts}

\newcommand{\ros}{{\it ROSAT}}

\newcommand{\xmm}{{\it XMM-Newton}}

\newcommand{\eros}{{eROSITA}} 

\newcommand{\srgl}{{\it Spectrum Roentgen Gamma}}
\newcommand{\srg}{{\it SRG}}
\newcommand{\nicer}{{\it NICER}}
\newcommand{\nicerl}{{Neutron star Interior Composition Explorer}}

\newcommand{\fermil}{{\it Fermi} Gamma-ray Space Telescope}
\newcommand{\fermi}{{\it Fermi}}
\newcommand{\fermilat}{{\it Fermi} Large Area Telescope}
\newcommand{\esol}{{European Southern Observatory Very Large Telescope}}
\newcommand{\eso}{{\it ESO-VLT}}
\newcommand{\lbtl}{{Large Binocular Telescope}}
\newcommand{\lbt}{{\it LBT}}

\newcommand{\nh}{N_{\rm H}}

\newcommand{\forstl}{{Focal Reducer/low dispersion Spectrograph 2}}
\newcommand{\forst}{{FORS2}}
\newcommand{\fastl}{{Five-hundred-meter Aperture Spherical radio Telescope}}
\newcommand{\fast}{{\it FAST}}

\def \calveraf{\object{1RXS~J141256.0+792204}}
\def \calvera{\object{Calvera}}
\def \zsfs{\object{J0657}}

\def \tmzsfs{\object{PSR~B0656+14}}
\def \tmgem{\object{Geminga}}
\def \tmozff{\object{PSR~B1055--52}}

\def \fluxcgs{erg~s$^{-1}$~cm$^{-2}$}
\def \jotos{\object{eRASSU~J131716.9--402647}}
\def \jzsfs{\object{eRASSU~J065715.3+260428}}
\def \jzsfsatnf{\object{PSR~J0657+2604}}

\begin{document} 
\title{
A multi-wavelength view of the isolated neutron star eRASSU~J065715.3+260428\thanks{Based on observations obtained with \xmm, an ESA science mission with instruments and contributions directly funded by ESA Member States and NASA (observation 0921280201)}}
\author{J.~Kurpas\inst{1,2}
\and A.~M.~Pires\inst{3,1}
\and A.~D.~Schwope\inst{1}
\and Z.~C.~Pan\inst{4,5,6,7}
\and Z.~L.~Zhang\inst{8,7}
\and L.~Qian\inst{5,4,6,9}
\and F.~Haberl\inst{10}
\and L.~Ji\inst{11}
\and I.~Traulsen\inst{1}}
\offprints{J. Kurpas}
\institute{Leibniz-Institut f\"ur Astrophysik Potsdam (AIP), An der Sternwarte 16, 14482 Potsdam, Germany
\email{jkurpas@aip.de} 
\and
Potsdam University, Institute for Physics and Astronomy, Karl-Liebknecht-Stra\ss e 24/25, 14476 Potsdam, Germany
\and
Center for Lunar and Planetary Sciences, Institute of Geochemistry, Chinese Academy of Sciences, 99 West Lincheng Rd., 550051 Guiyang, China
\email{adriana@mail.gyig.ac.cn}
\and
National Astronomical Observatories, Chinese Academy of Sciences, 20A Datun Rd., Chaoyang District, Beijing, 100101, People’s Republic of China
\and
Guizhou Radio Astronomical Observatory, Guizhou University, Guiyang, People’s Republic of China
\and
College of Astronomy and Space Sciences, University of Chinese Academy of Sciences, Chinese Academy of Sciences, Beijing, People’s Republic of China
\and
Key Laboratory of Radio Astronomy and Technology, Chinese Academy of Sciences, Beijing, People’s Republic of China
\and
Shanghai Astronomical Observatory, Chinese Academy of Sciences, Shanghai 200030, China
\and
CAS Key Laboratory of FAST, National Astronomical Observatories, Chinese Academy of Sciences, Beijing, 100101, China
\and
Max-Planck-Institut f\"ur extraterrestrische Physik, Gie{\ss}enbachstra\ss e 1, 85748 Garching, Germany
\and
Purple Mountain Observatory, Chinese Academy of Sciences, 10 Yuanhua Road, Nanjing 210023, China
}
\date{Received ...; accepted ...}
\titlerunning{Multi-wavelength study of \zsfs}
\authorrunning{J.~Kurpas et al.}
\abstract
{
On the premise of a soft spectral distribution and absence of counterparts, the X-ray source \jzsfs\ was recently identified as a likely thermally emitting isolated neutron star (XINS) in a search in the \srg/eROSITA All-Sky Survey. We investigated the nature and evolutionary state of the neutron star through a dedicated multi-wavelength follow-up observational campaign with \xmm, \nicer, \fast, and \eso, complemented by the analysis of archival \fermi-\textit{LAT} observations. The coherent timing analysis of the X-ray observations unveiled the rotation period of the XINS, $P=261.085400(4)$\,ms, and its spin-down rate, $\dot{P}=6^{+11}_{-4}\times10^{-15}$\,s\,s$^{-1}$ (errors are $1\sigma$ confidence levels). The nearly sinusoidal pulse profile has a pulsed fraction of $\sim$15\,\% ($0.2-2$\,keV). No optical counterparts are detected down to 27.3\,mag ($5\sigma$, $R$ band) in the \eso\ FORS2 imaging, implying a large X-ray-to-optical flux ratio above 5200. The X-ray spectrum of the source is best described by a composite phenomenological model consisting of two thermal components, either a double blackbody continuum with temperatures 90\,eV and 220\,eV or a hydrogen neutron star atmosphere of temperature $\log(T/\mathrm{K})\sim 5.8$ combined with a hot blackbody of 250\,eV, in both cases modified by an absorption feature at low energies, $\sim$0.3\,keV with an equivalent width of $\sim$100\,eV. The presence of faint non-thermal hard X-ray tails is ruled out above $(2.1\pm1.8)$\,\% of the source unabsorbed flux. 
Radio searches at $1-1.5$\,GHz with \fast\ yielded negative results, with a deep upper limit on the pulsed flux of 1.4\,$\mu$Jy ($10\sigma$). Similarly, no significant spatial or pulsed signals were detected in sixteen years of \fermi-\textit{LAT} observations. The most likely interpretation is that the source is a middle-aged spin-powered pulsar, which can also be identified as \jzsfsatnf. The absence of non-thermal X-ray, radio, or gamma-ray emission within current limits suggests either an unfavourable viewing geometry or unusual magnetospheric properties. Additional observations are needed to check for faint hard X-ray tails, investigate the presence of diffuse emission from a pulsar-wind nebula, and obtain a more accurately sampled timing solution.
}
\maketitle
\section{Introduction\label{sec_intro}}
In recent years, there has been renewed interest in identifying X-ray emitting isolated neutron stars (XINSs) that are not detected by conventional radio and gamma-ray pulsar surveys \citep{2022MNRAS.509.1217R,2022A&A...666A.148P,2024ApJ...961...36D,2023A&A...674A.155K,2024A&A...687A.251K}. While the detectability of old and distant radio pulsars may be affected by several selection biases, including beaming, nulling, and propagation effects in the interstellar medium (ISM), gamma-ray surveys such as those conducted by the \fermil\ \citep{2009ApJ...697.1071A} can miss pulsars if their gamma-ray emission is weak or if the pulsars are located in regions of strong gamma-ray background. X-ray observations, though affected by absorption at soft energies by the ISM, are sensitive to both thermal emission from the hot neutron star surface and the magnetospheric radiation powered by neutron star rotation. In particular, searches for thermally emitting XINSs provide a detection channel that is independent of the pulsar's spin and viewing geometry. 

Wide-field X-ray survey missions, such as \ros\ in the 1990s \citep{1982AdSpR...2d.241T} and \eros\ on board the recently launched \srgl\ (\srg) Observatory \citep{2021A&A...647A...1P}, greatly enhance our understanding of neutron star demographics, their physical properties, and their role in the broader astrophysical context. The X-ray source \jzsfs\ (hereafter dubbed \zsfs; \citealt{2023A&A...674A.155K,2024A&A...687A.251K}) is one of the first candidates identified in a search for XINS in the western Galactic hemisphere of the \eros\ All-Sky Survey \citep[eRASS;][]{2024A&A...682A..34M}. Its X-ray spectrum, best described by a single blackbody component with $kT\sim110$\,eV, is predominantly thermal and seemingly constant. Deep optical follow-up imaging with the \lbtl\ \citep{2012SPIE.8444E..1AH} in the $r$ band has excluded the presence of counterparts brighter than 25.5\,mag, suggesting a compact nature (X-ray-to-optical flux ratio $f_{\mathrm X}/f_{\mathrm opt}\gtrsim860$). 

Additional observations are needed to fully characterise the evolutionary state of the source and place it among the families of Galactic isolated neutron stars \citep[INSs; e.g.][for an overview]{2017AN....338..213P,2021ARep...65..615K,2023Univ....9..273P}. For example, the measurement of the INS rotation and spin evolution provides estimates of the magnetic field strength and characteristic age under the usual assumptions of a dipolar magnetic field configuration and loss of rotational energy solely due to magnetic braking in vacuum \citep{1969ApJ...157.1395O}. Spectral characterisation is equally important to constrain the presence of possible absorption features and non-thermal components. However, the shallow survey exposures of \eros\ (typically, $\sim150$\,s per sky visit for this source), and its comparably low effective area at energies above 1\,keV, prevent detailed timing and spectral analysis.

We report in this work the results of a dedicated multi-wavelength follow-up campaign targeting the XINS \zsfs. At X-ray energies, \zsfs\ was observed with the \nicerl\ \citep[\nicer;][]{2016SPIE.9905E..1HG} and \xmm\ \citep[][]{2001A&A...365L...1J}. In the optical, we obtained deep imaging with the \forstl\ (\forst) instrument at the \esol\ \citep[\eso;][]{1998Msngr..94....1A}. In the radio regime, we performed a DDT observation with the \fastl\ \citep[\fast;][]{2011IJMPD..20..989N}, which was carried out simultaneously with the \xmm\ pointing. We complemented the analysis with gamma-ray observations from the \fermilat. The XINS candidate \zsfs\ is the second \eros\ selected X-ray source to be the target of such a multi-wavelength effort. We previously reported the results of follow-up observations of \jotos, a long spin-period X-ray source displaying properties that closely resemble those of highly magnetised XINSs \citep{2024A&A...683A.164K}. These first promising results confirm \eros's ability to unveil elusive XINSs beyond the solar neighbourhood.

The outline of the paper is as follows: we begin with a description of the observations and data reduction in Sect.~\ref{sec_obs}. The results of the X-ray timing and spectral analysis, along with the search for optical, radio, and gamma-ray counterparts, are presented in Sect.~\ref{sec_analysis}. In Sect.~\ref{sec_disc}, we discuss the implications for the nature of \zsfs, particularly in the context of the known population of Galactic INSs. Finally, our conclusions are provided in Sect.~\ref{sec_concl}.
\begin{table}
\small
\caption{Overview of the X-ray observations\label{tab_obs}}
\centering
\scalebox{.91}{
\begin{tabular}{lccrr}
\hline\hline\noalign{\smallskip}
Observatory & Instrument & MJD\tablefootmark{(a)} & \multicolumn{1}{c}{GTI\tablefootmark{(b)}} & \multicolumn{1}{c}{Rate\tablefootmark{(c)}} \\
& & [days] & \multicolumn{1}{c}{[s]} & [$10^{-2}$\,cts\,s$^{-1}$]\\
\hline
\nicer & XTI & 60265.5432 & 116 038 &  90.3(4) \\
\xmm & EPIC pn & 60384.2996 & 44 954 & 15.04(21)\\
\xmm & EPIC MOS1 & 60384.6848 & 47 830 & 2.13(9)\\
\xmm & EPIC MOS2 & 60384.6847 & 49 733 & 2.46(9)\\
\hline
\end{tabular}}
\tablefoot{The programme identifiers of the \nicer\ and \xmm\ observations are 65720101 and 0921280, respectively. The EPIC cameras were operated in FF/SW imaging mode with thin filters.
\tablefoottext{a}{Modified Julian Date at mid-observation.}
\tablefoottext{b}{We list the remaining `good' observing time intervals after applying the data reduction steps and screening for periods of high background activity (see the text for details).}
\tablefoottext{c}{Observed count rate in the energy range $0.3-5$\,keV.}}
\end{table}
\section{Observations and data reduction\label{sec_obs}}
\subsection{\nicer\label{sec_obsnicer}}
Between November 9 and 25, 2023, \nicer\ observed the XINS candidate \zsfs\ continuously for approximately 210\,ks. We used the NICERDAS tools version 2024-02-09\_V012A, distributed as part of the HEASoft software release version 6.33.2, to process the observations. We applied the \texttt{nicerl2} task to extract the event lists and remove flagged events. We only considered intervals of orbital night-time to mitigate the impact of the optical leak affecting the scientific performance of \nicer\footnote{\url{https://heasarc.gsfc.nasa.gov/docs/nicer/analysis_threads/optical_leak}}. As \zsfs\ is a fairly faint and soft X-ray source, we adopted conservative screening criteria according to the \nicer\ analysis guidelines\footnote{Specifically, we set the \textit{underonly\_range} and \textit{overonly\_range} parameters to between 0 and 50, and 0 and 5, respectively. The \textit{cor\_range} parameter was set to be above 1.5. We removed `noise ringers' by setting the \textit{keep\_noisering} and the \textit{noisering\_under} options to `no' and 80. Lost events were removed setting the \textit{max\_lowmem} threshold to 250. Finally, to filter out events detected when \nicer\ was near the South Atlantic Anomaly, we set \textit{nicersaafilt} option to `no' and the \textit{saafilt} option to `yes'.}. The cleaned observations were merged by the \texttt{niobsmerge} task. Periods of remaining high flaring background were removed through a $3\sigma$-clipping algorithm from the mean count rate in the $0.3-12$\,keV energy band. The total net exposure after applying all screening criteria is 116\,ks and the total events $\sim8\times10^4$ ($0.3-2$\,keV; Table~\ref{tab_obs}). Spectra and light curves were extracted with the \texttt{nicerl3-spect} and \texttt{nicerl3-lc} pipelines. We estimated the background with the SCORPEON\footnote{\url{https://heasarc.gsfc.nasa.gov/docs/nicer/analysis_threads/scorpeon-overview/}} model. For the timing analysis, the \texttt{barycorr} task, JPLEPH.405 ephemeris, and the improved X-ray sky position of the target derived from the \xmm\ EPIC observation (Sect.~\ref{sec_obsxmm}) were used to correct the times-of-arrival (ToAs) to the barycentre of the Solar System.
\subsection{\xmm\label{sec_obsxmm}}
\begin{table*}
\caption{Results of source detection and astrometry\label{tab_pos}}
\centering
\scalebox{.98}{
\begin{tabular}{ccccccccc}
\hline\hline
RA & DEC & Pos.~Error\tablefootmark{(a)} & $l$\tablefootmark{(b)} & $b$\tablefootmark{(b)} & Ref.~Matches\tablefootmark{(c)} & Offset in RA & Offset in Dec & Likelihood\tablefootmark{(d)} \\
(degree) & (degree) & (arcsec) & (degree) & (degree) & & (arcsec) & (arcsec) & \\
\hline
104.313838 & 26.07500 & 0.20 & 189.95 & 12.75 & 27 & -1.38(18) & 0.04(20) & 9\\
\hline
\end{tabular}}
\tablefoot{
\tablefoottext{a}{We give the value for the $1\sigma$ confidence radius.}
\tablefoottext{b}{Galactic coordinates.}
\tablefoottext{c}{Number of matches between the EPIC and GSC source lists considered by \texttt{eposcorr}.}
\tablefoottext{d}{Likelihood of the boresight correction obtained with \texttt{eposcorr}.}}
\end{table*}
As part of large programme 092128, the \xmm\ Observatory targeted the XINS \zsfs\ on March 14/15, 2024, for $\sim$67\,ks (Table~\ref{tab_obs}). The EPIC pn detector \citep{2001A&A...365L..18S} was operated in full-frame (FF) mode, whereas the small-window (SW) mode was selected for EPIC MOS \citep{2001A&A...365L..27T}. Both cameras were equipped with the thin filter for its enhanced response at soft X-ray energies. We used the \xmm\ Science Analysis Software (version: 21.0.0) to process the observations. We found the observation to be significantly affected by periods of high background (about $30$\,\% of the elapsed time). Similarly to \nicer\ (Sect.~\ref{sec_obsnicer}), we applied a $3\sigma$-clipping algorithm to remove time intervals of flaring background. The total net exposure is $\sim45-50$\,ks for the individual EPIC cameras, amounting to a total $1.0\times10^4$ counts ($0.2-2$\,keV; Table~\ref{tab_obs}).

We conducted source detection with the \texttt{edetect\_stack} task across the five standard \xmm\ bands and three EPIC cameras \citep{2020A&A...641A.137T}. The astrometry was refined with the SAS task \texttt{eposcorr}, considering 27 potential matches of common sources from the optical Guide Star Catalogue \citep[version: 2.4.2;][]{2008AJ....136..735L} that are located within 15\arcmin\ of the nominal pointing coordinates of the EPIC observation. The resulting X-ray sky position is given in equatorial and Galactic coordinates in Table~\ref{tab_pos}. Based on the signal-to-noise ($S/N$) ratio, optimal source and background regions were defined with the SAS task \texttt{eregionanalyse} in the $0.2-12$\,keV energy band.

We followed the general SAS guidelines to extract light curves and spectra. Single and double pattern events were considered for pn (PATTERN $<=4$) while all valid patterns were included for the MOS detectors (PATTERN $<=12$). The extracted spectra were grouped using a maximum oversampling factor of three and at least 25 counts per energy bin. The barycentric correction was performed with the SAS \texttt{barycen} task, DE405 ephemeris, and the updated X-ray source position of the target listed in Table~\ref{tab_pos}.
\subsection{\eso}
The field of \zsfs\ was observed with the FORS2 instrument at the \eso\ in December 2023 for 1\,hour. In total, nine images in the $R_\mathrm{SPECIAL}$ band were taken, each with 400\,s exposure time. We used the EsoReflex environment with the FORS workflow for imaging data (version 5.6.2) to conduct the basic data reduction \citep{2013A&A...559A..96F}. The resulting images were subsequently astrometrically calibrated using the \texttt{astrometry.net} code \citep{2010AJ....139.1782L}. As several bright stars were saturated in the FORS images, an individual index file based on Legacy Survey DR10 \citep{2019AJ....157..168D} positions was adopted for this task. To facilitate stacking, the individual exposures were reprojected with the astropy \texttt{reproject} package \citep{astropy:2013, astropy:2018, astropy:2022}. Finally, we removed cosmic rays with the \texttt{L.A.Cosmic} algorithm \citep{2001PASP..113.1420V,curtis_mccully_2018_1482019} implemented in the \texttt{ccdproc} package \citep{matt_craig_2017_1069648}.
\subsection{\fast\label{sec_obsfast}}
To detect potential periodic signals in the radio band for \zsfs, \fast\ was used with the L-band 19-beam receiver \citep{2020RAA....20...64J} for the observations. 
The system temperature is approximately 24\,K and the central beam size at 1.4\,GHz is 2.9\arcmin. Given a pointing error of less than 8\arcsec\ and the target \zsfs\ having a position error of 0.2\arcsec\ (see Table \ref{tab_pos}), it should be well within the beam and very close to the centre. Observations were conducted on March 14, 2024, for a duration of 4.86 hours. Data were sampled with 8-bit precision for four polarisations, with a sampling time of 98.304\,$\mu$s and divided into 1024 channels covering 1 to 1.5\,GHz (with a channel width of 488\,kHz). The data were recorded in \texttt{PSRFITS} format \citep{2004PASA...21..302H} for subsequent analysis.
\section{Results\label{sec_analysis}}
\subsection{X-ray timing analysis\label{sec_xraytiming}}
\begin{figure}[t]
\begin{center}
\includegraphics[width=\linewidth]{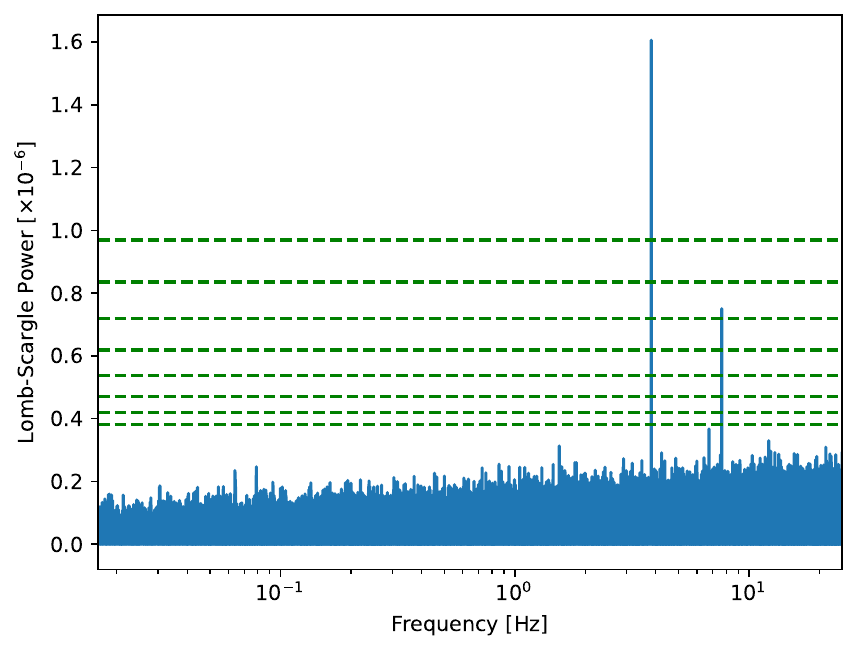}
\end{center}
\caption{Lomb-Scargle periodogram for the \nicer\ observation. The horizontal green dashed lines show the power corresponding to significance levels of $1\sigma$ to $8\sigma$.}
\label{fig_ls_periodogram}
\end{figure}

We searched for periodic modulations in the \nicer\ and \xmm\ EPIC pn datasets applying a Lomb-Scargle algorithm \citep{1976Ap&SS..39..447L,1982ApJ...263..835S}. We adopted time bins of 1\,ms for \nicer\ and 90\,ms for the \xmm\ light curves. We found both time series to contain a significant ($>4\sigma-8\sigma$) periodic signal at $\sim$261\,ms (see Fig.~\ref{fig_ls_periodogram} and Table~\ref{tab_tres}). While the first harmonic is detected at twice the fundamental frequency in the \nicer\ dataset, it cannot be detected in the \xmm\ data. Only a significant signal at 137\,ms is identified, that arises due to a beating between the 90\,ms binning and the 261\,ms modulation of the pulsar. We searched for shorter period modulations in the \nicer\ dataset (down to 1\,ms), but found no additional significant signal. The EPIC MOS observations were not included in the search, due to their insufficient time resolution\footnote{The time resolution is 300\,ms in SW mode (\url{https://xmm-tools.cosmos.esa.int/external/xmm_user_support/documentation/uhb/epicmode.html}).}.

\begin{figure}[t]
\begin{center}
\includegraphics[width=\linewidth]{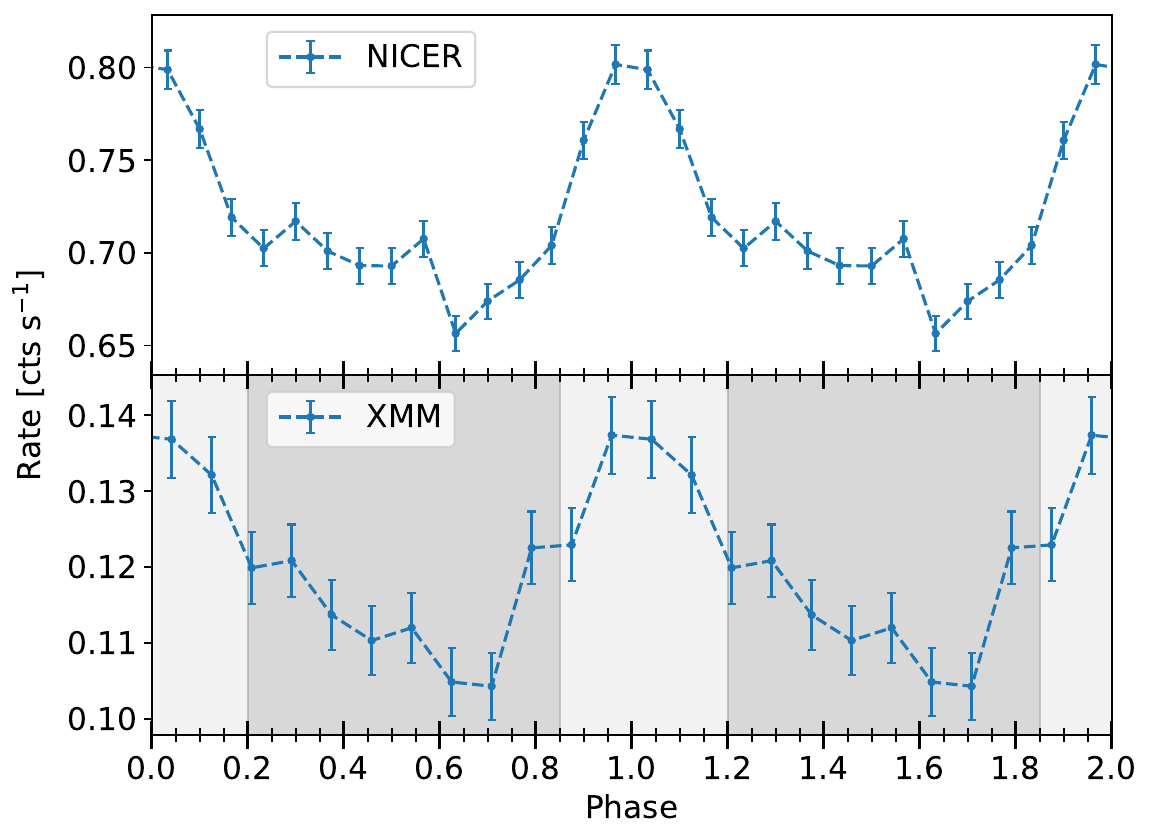}
\end{center}
\caption{X-ray pulse profiles from \nicer\ (\textit{top}) and \xmm\ (\textit{top}), folded at the 261\,ms modulation (see Table~\ref{tab_tres} and the text, for details). The two phase bins that were defined for the phase-resolved analysis of the \xmm\ data are marked by the light- and dark-grey regions in the lower panel.}
\label{fig_pulse_profile}
\end{figure}

The pulse profile in Fig.~\ref{fig_pulse_profile}, folded at the fundamental peak at 261\,ms, shows a single pulse and a saddle between phases 0.2 and 0.6 of the pulsar rotation. Folding the data at the first harmonics instead, leaves only a single broad peak for roughly half the pulsar rotation that can generally be well modelled with a single sinusoidal function (reduced chi-squared $\chi^2_\nu\sim1$ for $\nu=12$ degrees of freedom in the \nicer\ dataset). Since multiple harmonics are necessary to accurately model the pulse profile in Fig.~\ref{fig_pulse_profile}, we interpret the 261\,ms modulation as representing the true rotational period of \zsfs. This is because it includes more distinct and well defined features.

\begin{table*}
\small
\caption{X-ray timing ephemerides for the XINS \zsfs\ (\jzsfsatnf)\label{tab_tres}}
\centering
\scalebox{1.}{
\begin{tabular}{lrrr}
\hline\hline
& \nicer & \xmm & \nicer\ + \xmm\\
\hline
Counts & 80 051 & 7 638 & 87\ 689\\
Energy range [keV] & $0.3-2$ & $0.2-2$ & $0.2/0.3-2$\\
$\Delta T$\tablefootmark{(a)} [s] & 1\ 356\ 244 & 51\ 497& 10\ 980\ 326\\
$\nu_\mathrm{LS,min}-\nu_\mathrm{LS,max}$ [Hz] & $0.0167-25$ & $0.0167-10$\\
LS significance\tablefootmark{(b)} & $>8\sigma$ & $4.3\sigma$ & \\
$\Delta\nu_\mathrm{GL}$\tablefootmark{(c)} ($\mu$Hz)& 1 & 20 & \\
$O^\star_\mathrm{per}$ & $5\times10^{43}$ & $10^{15}$ & \\
$p_{\nu_1,\nu_2}$\tablefootmark{(d)} [\%] & $\sim$100 & $\sim$100 & \\
Frequency [Hz] & 3.830164336(24) & 3.8301649(17) & 3.83016439(6)\\
Period [ms] & 261.0854032(16) & 261.08536(12) & 261.085400(4)\\
Period derivative [s\ s$^{-1}$] & & & $6^{+11}_{-4}\times10^{-15}$\\
Reference time\tablefootmark{(e)} [MJD] & 60265.4441739 & 60384.3138306 & 60265.4441739\\
Pulsed fraction\tablefootmark{(f)} [\%] & $10.0\pm1.0$ & $13.7\pm2.8$ & \\
Dipolar magnetic field [G] & & & $1.2^{+0.9}_{-0.6} \times 10^{12}$\\
Characteristic age [Myr] & & & $0.7^{+1.7}_{-0.5}$\\
Spin-down luminosity [erg\,s$^{-1}$] & & & $2.0^{+4}_{-1.5} \times 10^{34}$\\
\hline
\end{tabular}}
\tablefoot{Event arrival times used in the timing analysis were baricentrically corrected using the updated X-ray sky position of $104.313838\deg$/$26.07500\deg$ and the JPL~DE405 solar system ephemeris.
\tablefoottext{a}{Elapsed time between the first and last photon of the observation.}
\tablefoottext{b}{Frequency step used in the \citet[][]{1996ApJ...473.1059G} search.}
\tablefoottext{c}{For the fundamental peak, we show the significance that the observed Lomb-Scargle power is not due to white noise. The chance that the observed peak is due to noise was computed from the false-alarm probability \citep{2008MNRAS.385.1279B}.}
\tablefoottext{d}{Probability that the probed frequency interval contains a periodic signal \citep[see][for details]{2000ApJ...540L..25Z}.}
\tablefoottext{e}{Arbitrarily defined reference time in MJD (TDB) that aligns the \nicer\ and \xmm\ pulse profiles (Fig.~\ref{fig_pulse_profile}). The reference time is chosen in such a way that phase-zero roughly coincides with the centre of the pulse.}
\tablefoottext{f}{Relative fraction between the minimum and maximum count rate in the pulse profile in Fig.~\ref{fig_pulse_profile}, computed via $p_{\rm f} = \frac{R_\mathrm{max}-R_\mathrm{min}}{R_\mathrm{max}+R_\mathrm{min}}$ in the $0.2-2$\,keV band.}}
\end{table*}

With the goal to refine the period estimation for a phase-coherent analysis, we applied the \citet[][]{1996ApJ...473.1059G} method \citep[see e.g.][for a description]{2024A&A...683A.164K}. Specifically, we adopted $m_{max} = 12$ and frequency steps ($\Delta\nu_\mathrm{GL}$) of 1\,$\mu$Hz and 20\,$\mu$Hz for \nicer\ and \xmm, respectively. The results are shown in Table~\ref{tab_tres}. The spin period and pulsed fraction between \xmm\ and \nicer\ agree within $2\sigma$; we note that the deviations between the datasets may be attributed to the stronger \nicer\ background and the higher timing accuracy in comparison to EPIC pn. 

The mid-observation times of \nicer\ and \xmm\ are roughly separated by 120 days. The linear timing solution from \nicer\ (arbitrarily centred at the phase of the maximum of the pulse profile, see Table~\ref{tab_tres}), indicates that both observations can be connected in phase with a cycle counting error of 0.25. However, folding the \xmm\ events with the \nicer\ timing solution reveals a phase shift of 0.4 between the maxima of the pulse profiles, indicating the presence of a measurable spin-down between the two datasets.

To find a best-fitting phase coherent timing solution, we defined the residuals between observed and tested cycle counting as
\begin{equation}
\label{eq_toares}
R = \sum_{i=1}^{N} \left(\frac{C_\mathrm{obs,i} - C(t_i,\nu,\dot{\nu},t_0,C_0)}{\sigma(C_\mathrm{obs,i})}\right)^2,
\end{equation}
\noindent where the summation is over the $N$ defined ToAs of the events.
In Eq.~(\ref{eq_toares}), we adopted a quadratic timing solution of the form:
\begin{displaymath}
C(t,\nu,\dot{\nu},t_0,C_0) = C_0+\nu(t-t_0)+\frac{1}{2}\dot{\nu}(t-t_0)^2,
\end{displaymath}  
\noindent where $\nu$ and $\dot{\nu}$ is the tested spin and spin-down, $t_0$ is the reference time, and $C_0$ is the phase shift.

We split the \nicer\ exposure into sections of at least 13\ 000 photons, resulting in a series of six ToAs where the XINS pulsation could be reliably detected. 
To determine the phase of each pulse, we fitted the pulse profile (folded with the tested timing solution) with a sinusoidal function including the fundamental and first harmonic. The fit parameters were then used to estimate the phase of the maximum of the pulse profile and, consequently, the ToA that is closest to midpoint of each time section. We then searched for the timing solution that minimised the cycle counting residuals for our data.

We explored the likelihood landscape of possible timing solutions through the nested sampling algorithm implemented in the \texttt{UltraNest} package \citep{2021JOSS....6.3001B}. We assumed linear priors for the period and phase shift, whereas a logarithmic prior was used for the spin-down. The reference time was kept near the mid-observation of the \nicer\ dataset and near the time of phase zero at $t_0=60265.4441739$. In the analysis, we evolved 800 sample points until the remaining unexplored fraction of the evidence integral amounted to less than 1\%.

\begin{figure}[t]
\begin{center}
\includegraphics[width=\linewidth]{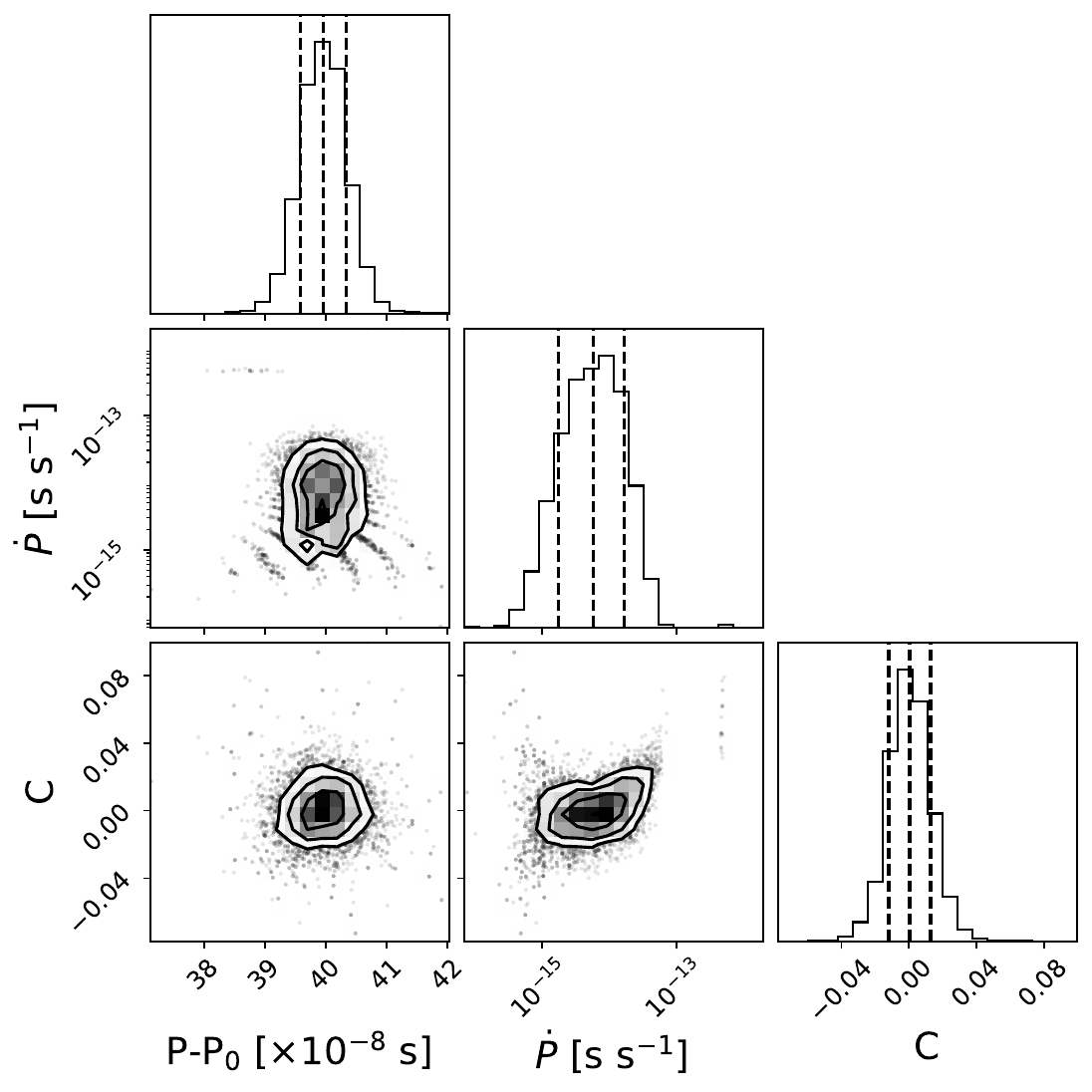}
\end{center}
\caption{Corner plot depicting the parameter distribution of the coherent timing search. For presentation purposes, we subtracted the period $P_0=261.085$\,ms from the period values of the distribution. Vertical lines mark the 15.8\%, median and 84.1\% percentiles.}
\label{fig_tsol_prob_distr}
\end{figure}

The resulting posterior probability distribution is shown in Fig.~\ref{fig_tsol_prob_distr}. We obtained from the distribution the best solution $P = 261.085400(4)$\,ms, $\dot{P} = 6^{+11}_{-4}\times10^{-15}$\,s\,s$^{-1}$, and $C_0 =  0.001^{+0.013}_{-0.012}$ (we give the median value and confidence range, containing 68\% of the sample points). We note, however, that this timing solution still allows for ambiguity in the cycle counting. While the solution constrains the possible range of values for $P$ and $\dot{P}$, there are still multiple points within the confidence region space that can minimise the cycle counting residuals, but may deviate by multiple cycles from each other. Additional observations covering a wider sampling in time will be needed to correctly identify the best-fit solution from these low residual points.

\subsection{X-ray spectral analysis\label{sec_x_ray_fits}}
\subsubsection{Phase-integrated analysis}
The spectral analysis of the \xmm\ and \nicer\ datasets was performed with \texttt{XSPEC} \citep[version: 12.14.0h,][]{1996ASPC..101...17A}. All models were coupled with a \texttt{tbabs} component to account for interstellar absorption \citep[adopting elemental abundances from ][]{2000ApJ...542..914W}. While analysing the spectra from multiple instruments simultaneously (XTI, EPIC pn, and MOS1/2), we included an additional constant factor in the fit to address cross-calibration uncertainties. Chi-squared statistics were employed to fit the spectral data. 

The results are presented in Table~\ref{tab_fitres}. We initiated the spectral analysis using the EPIC pn dataset for simplicity. Initially, a single-component absorbed blackbody (\texttt{BB}) model did not provide a satisfactory fit ($\chi^2_\nu(\nu) = 5.29(42)$; see fit I in Table~\ref{tab_fitres}). Including a second \texttt{BB} component improved the fit residuals compared to fit I (cf.~fit II in Table~\ref{tab_fitres}). Alternatively, a simple absorbed power-law model (\texttt{PL}; fit IV in Table~\ref{tab_fitres}) closely matched the data. However, in this scenario, the best-fit column density $\nh$ exceeds the Galactic value along the line of sight, $N_{\rm H,gal}=8.27\times10^{20}$\,cm$^{-2}$  \citep{2016A&A...594A.116H}, and the photon index $\alpha$ of the power-law model is much steeper than typically observed in rotation-powered pulsars \citep[RPPs;][]{2009ASSL..357...91B}.
\begin{table*}
\caption{Results of the phase-integrated X-ray spectral analysis
\label{tab_fitres}}
\centering
\scalebox{.84}{
\begin{tabular}{cccccccccccccc}
\hline\hline
\multicolumn{5}{l}{EPIC pn (\texttt{BB}, \texttt{PL})}\\
\hline
& $\nh$ & $kT_1$ & $R_1$\tablefootmark{(a)} & $kT_2$ & $R_2$\tablefootmark{(a)} &$\alpha$ & $\epsilon$\tablefootmark{(b)} & $\sigma$ & $EW$\tablefootmark{(c)} & $\chi^2_\nu(\nu)$ & Absorbed flux\tablefootmark{(d)}\\
& $[10^{20}$\,cm$^{-2} ] $ & [eV] & [km] & [eV] & [km] & & [eV] & [km]/[eV] & [eV] & & [$10^{-13}$\,\fluxcgs]\\
\hline\noalign{\smallskip}
I  & $1.8^{+0.5}_{-0.5}$  &  $122.3^{+2.3}_{-2.2}$  &  $1.2^{+0.08}_{-0.08}$  &  &  &  &  &  &  & 5.29(42) &  $2.44^{+0.04}_{-0.04}$ \\
II  & $7.1^{+1.1}_{-1.0}$  &  $88^{+4}_{-4}$  &  $3.3^{+0.8}_{-0.5}$  &  $219^{+15}_{-13}$  &  $0.18^{+0.05}_{-0.04}$  &  &  &  &  & 1.22(40) &  $2.41^{+0.03}_{-0.03}$ \\
III  & $<5$  &  $93^{+6}_{-7}$  &  $2.2^{+0.6}_{-0.4}$  &  $222^{+17}_{-15}$  &  $0.16^{+0.05}_{-0.04}$  &  &  $306^{+50}_{-24}$  &  $51^{+19}_{-40}$  & $120^{+80}_{-100}$  & 1.08(37) &  $2.57^{+0.04}_{-0.04}$ \\
IV  & $15.7^{+0.9}_{-0.9}$  &  &  &  &  &  $5.05_{-0.10}^{+0.10}$  &  &  &  & 1.35(42) &  $2.377^{+0.029}_{-0.029}$ \\
V  & $12.8^{+1.7}_{-1.6}$  &  $88^{+9}_{-9}$  &  $2.1^{+1.0}_{-0.7}$  &  &  &  $4.63^{+0.24}_{-0.25}$  &  &  &  & 1.32(40) &  $2.41^{+0.03}_{-0.03}$ \\
VI  & $9.7^{+2.4}_{-4}$  &  &  &  &  &  $4.77^{+0.15}_{-0.18}$  &  $350^{+40}_{-50}$  &  $53^{+29}_{-30}$  &  $80^{+100}_{-50}$  & 1.06(39) &  $2.47^{+0.04}_{-0.04}$ \\
VII  & $<11$  &  $170^{+50}_{-70}$  &  $0.2^{+1.1}_{-0.11}$  &  &  &  $4.6^{+0.5}_{-0.4}$  &  $290^{+70}_{-70}$  &  $80^{+40}_{-40}$  &  $200^{+200}_{-140}$  & 1.08(37) &  $2.55^{+0.04}_{-0.04}$ \\
\noalign{\smallskip}\hline\noalign{\smallskip}
\multicolumn{5}{l}{EPIC pn (\texttt{NSA})\tablefootmark{(e)}}\\
\hline\noalign{\smallskip}
& $\nh$ & $\log(T)$ & $d$ & $kT$ & $R_{\rm BB}$\tablefootmark{(f)} & $\alpha$ & $\epsilon_1$ & $\sigma_1$ & $EW$\tablefootmark{(b)} & $\chi^2_\nu(\nu)$ & Absorbed flux\tablefootmark{(c)}\\
& [$10^{20}$\,cm$^{-2}$] & [$\log(\mathrm{K})$] & [kpc] & [eV] & [km] &  & [eV] & [eV] & [eV] & & [$10^{-13}$\,\fluxcgs]\\
\hline\noalign{\smallskip}
VIII & ${5.2}_{-0.6}^{+0.6}$ & ${5.873}_{-0.014}^{+0.013}$ & ${1.56}_{-0.16}^{+0.16}$ & & & & & & & 3.17(42) & ${2.41}_{-0.03}^{+0.03}$\\
IX & $<11$ & ${6.04}_{-0.23}^{+0.14}$ & ${11}_{-7}^{+7}$ & & & ${4.6}_{-0.5}^{+0.6}$ & ${280}_{-60}^{+70}$ & ${80}_{-40}^{+29}$ & $220^{+130}_{-150}$ & 1.07(37) & ${2.58}_{-0.04}^{+0.04}$\\
X & $<8$ & ${5.75}_{-0.06}^{+0.07}$ & ${0.7}_{-0.4}^{+0.4}$ & ${246}_{-22}^{+40}$ & ${0.08}_{-0.06}^{+0.12}$ & & ${320}_{-50}^{+50}$ & ${50}_{-40}^{+25}$ & $110^{+110}_{-80}$ & 1.04(37) & ${2.52}_{-0.04}^{+0.04}$\\
\noalign{\smallskip}\hline
\multicolumn{5}{l}{All instruments}\\
\hline
& $\nh$ & $kT_1$ & $R_1$\tablefootmark{(a)} & $kT_2$ & $R_2$\tablefootmark{(a)} & & $\epsilon$ & $\sigma$ & $EW$\tablefootmark{(b)} & $\chi^2_\nu(\nu)$ & Absorbed flux\tablefootmark{(c)}\\
& $[10^{20}$\,cm$^{-2} ] $ & [eV] & [km] & [eV] & [km] & & [eV] & [eV] & [eV] & & [$10^{-13}$\,\fluxcgs]\\
\hline\noalign{\smallskip}
XI & $<4$  &  $97^{+4}_{-6}$  &  $1.97^{+0.8}_{-0.028}$  &  $228^{+15}_{-14}$  &  $0.148^{+0.04}_{-0.027}$  & &   $289^{+50}_{-25}$  &  $58^{+17}_{-27}$  &  $130^{+70}_{-90}$  &  1.02(113)  &  $2.550^{+0.028}_{-0.028}$ \\
XII & $<1.1$ & $101^{+4}_{-5}$ & $1.92^{+0.21}_{-0.13}$ & $247^{+19}_{-18}$ & $0.122^{+0.04}_{-0.025}$ & & $285^{+17}_{-50}$ & $60^{+26}_{-17}$ & $130^{+100}_{-50}$ & 1.02(1086) & $2.696^{+0.022}_{-0.022}$\\
\noalign{\smallskip}\hline
\end{tabular}}
\tablefoot{We provide $1\sigma$ confidence intervals on the best-fit parameters.
\tablefoottext{a}{For the blackbody emission radius at infinity, we assumed a source at a distance of 1\,kpc.}
\tablefoottext{b}{Central line energy of the Gaussian absorption component.}
\tablefoottext{c}{The equivalent width (EW) is estimated from $\int \frac{f_c-f_o}{f_c} dE$, with $f_c$ being the continuum and $f_o$ the observed flux. Errors provide the maximum and minimum EW values obtained from all possible combinations of the upper and lower $1\sigma$ confidence interval limits of the model parameters.}
\tablefoottext{d}{The absorbed model flux covers the $0.2-10$\,keV range.}
\tablefoottext{e}{The model parameters assume a canonical neutron star with 1.4\,M$_\odot$ and 12\,km radius and a magnetic field strength of $10^{12}$\,G.}
\tablefoottext{f}{The size of the emission region of the blackbody component in combination with the \texttt{NSA} model is computed for a source at the best-fit distance 700\,pc.}}
\end{table*}

\begin{figure}[t]
\begin{center}
\includegraphics[width=\linewidth]{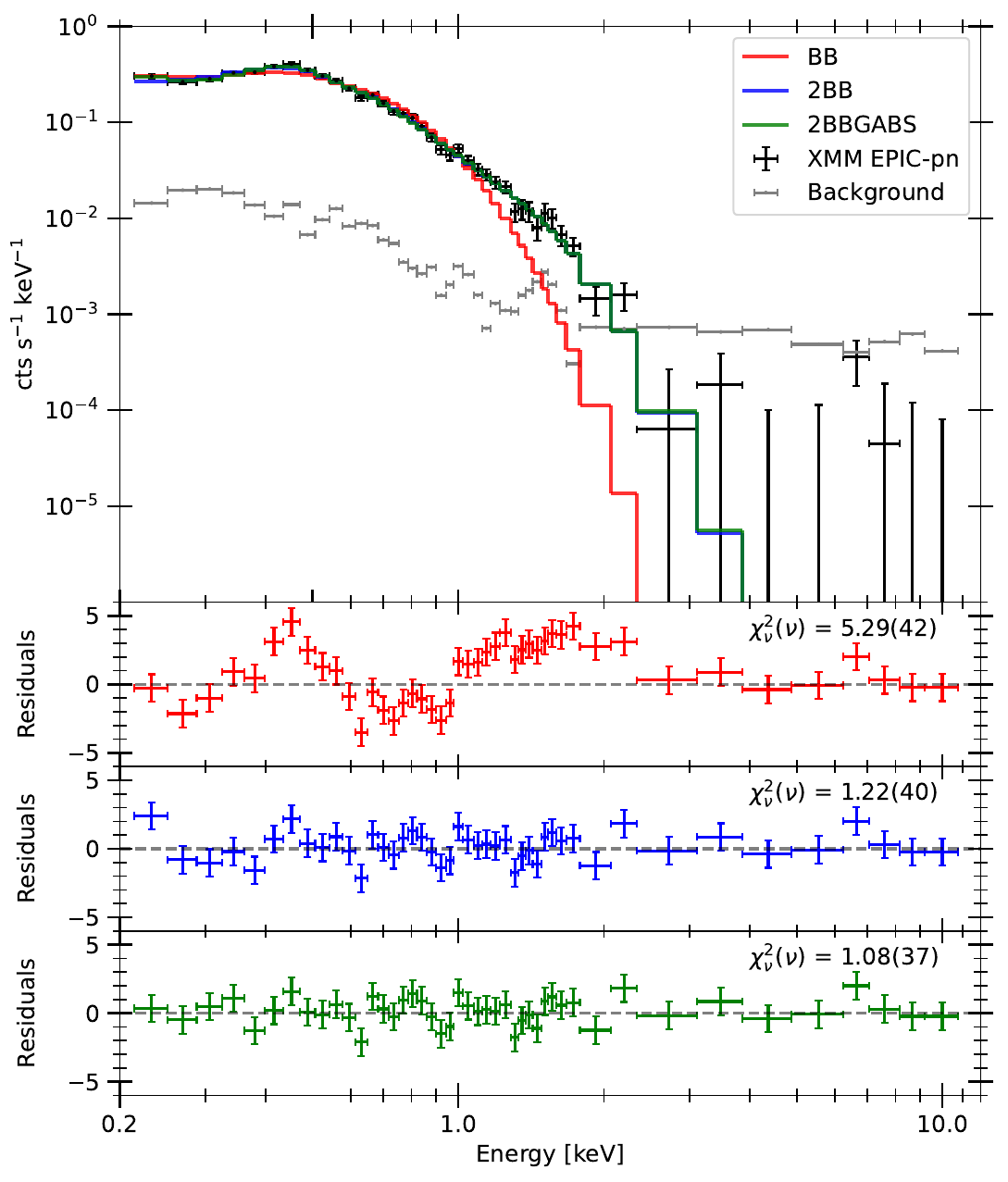}
\end{center}
\caption{EPIC pn spectrum of \zsfs\ along with the best-fit single blackbody (red, I), two blackbody (blue, II) or two blackbody with a line (green, III) models. See Table~\ref{tab_fitres} for the fit parameters.}
\label{fig_spec_plots}
\end{figure}

Absorption features have been observed in the spectra of several XINSs \citep[e.g.][]{2013Natur.500..312T, 2022A&A...661A..41S,2023IAUS..363..288M}. Similarly, for \zsfs, we found that including a Gaussian absorption component (\texttt{GABS}) at low energies ($\sim$0.3\,keV) significantly enhances the fit quality (models III and VI in Table~\ref{tab_fitres}), although there exists a strong degeneracy between the line width $\sigma$ and $\nh$ for model III. No improvement was observed when $\sigma$ was held fixed during fitting. For model VI, the power-law spectral slope remains significantly steeper than what is typically observed in the known INS population. We present the folded spectrum in Fig.~\ref{fig_spec_plots} along with the best-fit \texttt{2BBGABS} model. For comparison, the fit residuals using single or double \texttt{BB} components are also shown. For an assumed distance of 1\,kpc, we derive the size of the emission region for the dominant colder \texttt{BB} component in models I, II, and III. With values ranging from 1 to 4\,km the emission region size is below the canonical neutron star radius of $\sim$12\,km, but still consistent with phenomenological models used to describe the spectra of the purely thermally emitting XINSs \citep[e.g.~the `magnificent seven' described by e.g.~][]{2009ASSL..357..141T,2020MNRAS.496.5052P}.

The fit of a composite model combining thermal and non-thermal components reveals a steep \texttt{PL} and a hot \texttt{BB} component (fit result VII in Table~\ref{tab_fitres}), contrasting with middle-aged RPPs where soft thermal components dominate \citep[e.g.~the `three musketeers' described by ][]{2005ApJ...623.1051D}. In these sources, X-ray emission is often interpreted as a blend of thermal radiation from the cooling surface and hot polar caps. Additionally, fainter power-law tails extending to higher energies are attributed to magnetospheric activity of the pulsar.

We calculated upper limits for the detection of comparable non-thermal components using a \texttt{2BBGABSPL} model. We held constant the parameters of the \texttt{BB} and \texttt{GABS} components at the optimal values from model III, while assuming a \texttt{PL} slope of $\alpha=2$. The normalisation of the power-law component was allowed to vary freely. This analysis yielded a $3\sigma$ upper limit of $2.1\times10^{-14}$\,erg\,s$^{-1}$\,cm$^{-2}$, equivalent to about $(2.1\pm1.8)\%$ of the source's unabsorbed flux of model III.

Qualitatively similar outcomes were observed using neutron star atmosphere \citep[\texttt{NSA}; ][]{1995ASIC..450...71P,1996A&A...315..141Z} models (fit results VIII, IX, and X in Table~\ref{tab_fitres}). A single \texttt{NSA} component alone fails to adequately fit the data, regardless of the magnetic field intensity. However, composite models incorporating the 0.3\,keV absorption feature, alongside either a hot \texttt{BB} component with $kT\sim250$\,eV or a steep \texttt{PL} with $\alpha\sim4.6$, achieve statistically satisfactory fits. The temperature and normalisation of the atmosphere are notably affected by the inclusion of additional components: in the \texttt{NSAGABSPL} scenario, we observe a hot atmosphere and a substantial distance, while in the \texttt{NSAGABSBB} scenario, a colder atmosphere and a closer source are inferred.

To assess the significance of the absorption feature, we calculated both the false-negative (where an existing feature is missed or incorrectly identified) and false-positive (where a non-existing feature is identified) rates. For the false-negative rate, we utilised the \texttt{XSPEC} \texttt{fakeit} command to simulate 1,000 spectra based on the best-fit result from Table~\ref{tab_fitres} (III), employing a \texttt{2BBGABS} model. We then performed fits using both a \texttt{2BB} model and the \texttt{2BBGABS} model to determine how often the simulated absorption line would be undetected. We found a low false-negative rate of only 6.2\%, indicating that the spectral parameters were generally well recovered and confirming that the EPIC pn observation accurately captures the source's spectral properties.

For the false-positive rate, we simulated 1,000 spectra using the best-fit solution from Table~\ref{tab_fitres} (II), employing a \texttt{2BB} model. These simulated spectra were then fitted using both a \texttt{2BB} model and a \texttt{2BBGABS} model. We found that in only 2.7\% of all simulations did the addition of the absorption line improve the fit statistic by more than the observed $\Delta \chi^2_\nu > 0.14$. This low false-positive rate suggests a high significance that the observed feature is real.

The data from additional X-ray instruments, EPIC MOS1/2 and NICER XTI, were next included in the analysis. Simultaneous fits with all EPIC instruments aligned with previous results (fit XI in Table~\ref{tab_fitres}). However, incorporating \nicer\ data was challenging, as the modelling requires correct handling of background contamination, especially if bright field sources are present in the vicinity of the target. In particular, we identified in the EPIC images an X-ray source of similar brightness to the XINS, $2.10(5)\times 10^{-13}$ \fluxcgs\ ($0.2-10$\,keV), 2\arcmin\ north-west of its position. Its EPIC pn spectrum is best-fit by an absorbed \texttt{PL} with $\nh = 7.20^{+1.1}_{-1.0} \times 10^{20}$\,cm$^{-2}$ and $\alpha=1.98(6)$. To address this contamination and the large background at soft energies, a power-law component was added to the spectral model, and only \nicer\ photons above 500\,eV were included. The final simultaneous spectral fit aligned well with EPIC pn results (XII), except for some flux discrepancies likely due to background contributions or calibration inaccuracies.
\subsubsection{Phase-resolved analysis}
\begin{table*}
\caption{Results of the phase-resolved X-ray spectral analysis
\label{tab_fitphres}}
\centering
\scalebox{1.}{
\begin{tabular}{llccccccccccccc}
\hline\hline\noalign{\smallskip}
Model & Phase & $\nh$ & $kT_1$ & $R_1$\tablefootmark{(a)} & $kT_2$ & $R_2$\tablefootmark{(a)} & $\epsilon$ & $\sigma$ & EW\tablefootmark{(b)} & $\chi^2_\nu (\nu)$\\
& & $[10^{20}$\,cm$^{-2} ] $ & [eV] & [km] & [eV] & [km]& [eV] & [eV] & [eV] &\\
\hline\noalign{\smallskip}
\texttt{2BB} & off-pulse & $7.3^{+1.1}_{-1.0}$ & $86^{+5}_{-5}$ & $3.3^{+0.8}_{-0.5}$ & $212^{+18}_{-16}$ & $0.18^{+0.06}_{-0.04}$ & & & & 0.80(60) \\
& pulse & & $92^{+5}_{-5}$ & $3.0^{+0.7}_{-0.4}$ & $244^{+23}_{-21}$ & $0.13^{+0.05}_{-0.03}$ \\ 
\texttt{2BBGABS} & off-pulse & $2.5^{+2.3}_{-2.2}$ & $87^{+8}_{-8}$ & $2.6^{+1.4}_{-0.8}$ & $211^{+21}_{-18}$ & $0.18^{+0.08}_{-0.05}$ & $324^{+28}_{-21}$ & $51$ & $90^{+40}_{-40}$ & 0.69(56)\\
& pulse & & $96^{+7}_{-7}$ & $2.3^{+0.9}_{-0.5}$ & $245^{+26}_{-23}$ & $0.13^{+0.06}_{-0.04}$ & $299^{+27}_{-29}$ & $51$ & $100^{+40}_{-50}$ \\
\noalign{\smallskip}\hline
\end{tabular}}
\tablefoot{We provide $1\sigma$ confidence intervals on the best-fit parameters.
\tablefoottext{a}{For the blackbody emission radius at infinity, we assumed a source at a distance of 1\,kpc.}
\tablefoottext{b}{EW estimated as described in the notes of Table~\ref{tab_fitres}.}}
\end{table*}
Achieving a detailed phase-resolved analysis with the current dataset is challenging. For EPIC pn, the frame time in FF mode is a significant fraction of the spin period of \zsfs, leading to some event confusion even with only three phase bins. Although \nicer\ data could, in principle, provide a more detailed phase-resolved analysis, the increased background and contamination led us to focus on the EPIC pn data.

Within these limitations, we defined two phase intervals aimed at distinguishing the pulsed ($\phi_{\rm pulse}=0.85-0.2$) and non-pulsed ($\phi_{\rm off-pulse}=0.2-0.85$) emission of \zsfs\ (see the grey background in the lower panel in Fig.~\ref{fig_pulse_profile}). For each phase bin, we extracted the source spectrum. We did not correct the exposure for the camera dead time\footnote{The \texttt{LIVETIME} parameter computed by the \xmm\ SAS was unreasonably short, leading to overestimated count rates.}.

We conducted simultaneous fits of the pulsed and non-pulsed spectra, fixing only the $\nh$ value and $\sigma$ width of the line and leaving other model components free. The results were similar to those for the phase-integrated spectrum, with \texttt{PL} components converging to unreasonably large slopes and requiring at least two \texttt{BB} components to fit the spectrum. For the double blackbody models in Table~\ref{tab_fitphres}, the additional \texttt{GABS} absorption component was not necessary, although it improved the fit slightly and led to smaller $\nh$ values. The model parameters showed a maximum 2$\sigma$ variation between phase bins, which is not significant. The absolute temperatures of both \texttt{BB} components seemed slightly higher during the pulse maximum. Additional observations are needed to confirm any phase-variable trends.
\subsection{Searches for pulsed radio emission\label{sec_pulsefast}}
The pulsar search, which included both single pulse search and periodic signal detection, was conducted using \texttt{PRESTO~V4.0}\footnote{\url{https://github.com/scottransom/presto}} \citep{2011ascl.soft07017R}. The Radio Frequency Interference (RFI) was firstly masked by the \texttt{rfifind} routine, with the search for RFI in the frequency domain conducted over a time span of 12.9 seconds. Then, the data were de-dispersed by the \texttt{prepsubband} routine, which were corrected for the dispersion effects based on the pulsar's position and distance (measured in kpc). We used the predicted dispersion measure (DM) from the model by \citet[][YMW16]{2017ApJ...835...29Y}. According to this model, the predicted DM is $<$200 pc\,cm$^{-3}$. To ensure we captured any potential signals, we set the upper limit of the DM range to twice the predicted value, which is $\sim$400 pc\,cm$^{-3}$, and the DM was incremented in steps of 0.1\,pc\,cm$^{-3}$ across the entire range from zero.

The de-dispersed time series were used for further single pulse searches, or transformed into the frequency domain via Fast Fourier Transform (FFT) for periodic signal detection. For the single pulse search, the \texttt{PRESTO} routine \texttt{single\_pulse\_search.py} was applied. The \texttt{accelsearch} routine was used to search for periodic signals. Although \zsfs\ does not appear to be a binary pulsar, we set the maximum allowable frequency shift in the frequency domain (measured in Fourier bins) to 50 using the `-zmax' option. The candidates were produced for further identification and confirmation by both the \texttt{ACCEL\_sift.py} routine from \texttt{PRESTO} and manual selection after determining the possible spinning periods. None of the candidates were identified as either \zsfs\ or any other pulsar.

The sensitivity of our search can be estimated using the radiometer equation \citep[e.g.][]{2011ApJ...734...89L}:
\begin{displaymath}
S_{\rm min} = \beta \frac{(S/N_{\rm min}) T_{\rm sys}}{G \sqrt{n_{\rm p} t_{\rm int} \Delta f}} \sqrt{\frac{W}{P - W}},
\end{displaymath}
where $\beta$ represents the sampling efficiency and is equal to 1 for our 8-bit recording system. In this equation, $W$ denotes the width of the pulsar profile; we used 10\% of the pulse width for our calculations. The minimum signal-to-noise ratio, $(S/N)_{\rm min}$, for our search is set to 10. The system temperature, $T_{\rm sys}$, is 24\,K, and the antenna gain, $G$, is 16\,K\,Jy$^{-1}$. The number of polarisations, $n_{\rm p}$, is 2. The integration time $t_{\rm int}$ is measured in seconds, and the bandwidth $\Delta f$ is 300\,MHz. This bandwidth accounts for the fact that approximately 25\% of the data were masked as RFI and thus excluded from further processing. Consequently, the radio L-band flux of \zsfs\ should be lower than $1.4$\,$\mu$Jy ($10\sigma$).
\subsection{Optical limit}
\begin{figure}[t]
\begin{center}
\includegraphics[width=\linewidth]{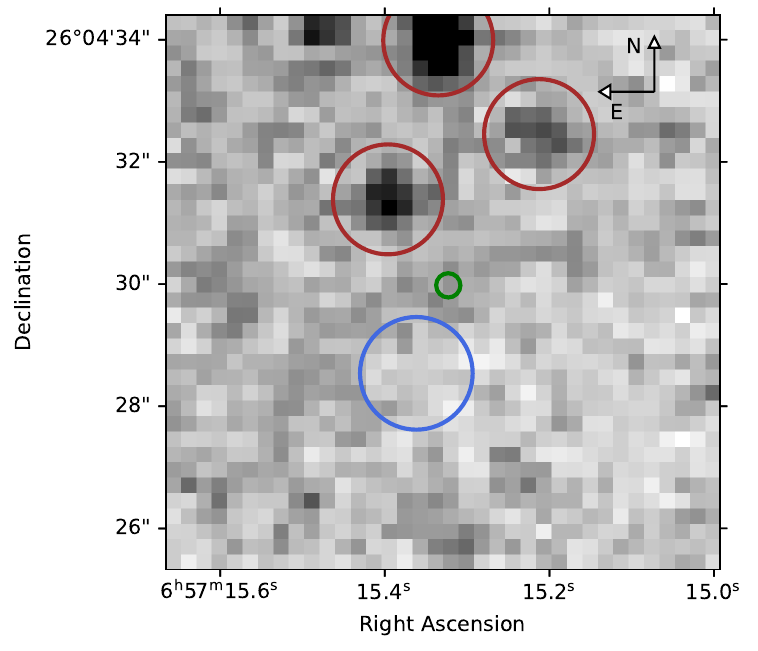}
\end{center}
\caption{FORS2 $R$-band image of the field of \zsfs. The \eros\ \citep{2023A&A...674A.155K,2024A&A...683A.164K} and \xmm\ X-ray sky positions are marked by the blue and green circles, respectively (radius marks the $1\sigma$ positional accuracy). Brown circles (with arbitrary radius of 1\arcsec) mark field sources identified from an \texttt{SExtractor} run.}
\label{fig_vlt_img}
\end{figure}

\begin{table}
\caption{Optical photometric parameters\label{tab_phot}}
\centering
\scalebox{1.}{
\begin{tabular}{lrccccc}
\hline\hline
Magnitude zero-point (ZP)\tablefootmark{(a)} [mag] & 28.30\\
Extinction (E)\tablefootmark{(a)} [mag] & 0.087\\
Airmass (AM) & 1.63\\
FWHM [pixel] & 2.44\\
FWHM [\arcsec] & 0.61\\
$\sigma_{\rm sky}$ & 0.15\\
Detection limit\tablefootmark{(b)} [$5\sigma$; mag] & 27.32\\
X-ray-to-optical flux ratio [$5\sigma$] & $>5260$\\
\hline
\end{tabular}}
\tablefoot{
\tablefoottext{a}{The nightly zero-point and extinction values were taken from the FORS Absolute Photometry project (\url{https://archive.eso.org/qc1/qc1_cgi?action=qc1_browse_table&table=fors2_photometry}).}
\tablefoottext{b}{The $5\sigma$ detection limit was calculated for a point source within an optimal Gaussian aperture via $m_\mathrm{lim} = ZP-2.5\log\left(6.75\times \sqrt{\pi}\times \frac{FWHM}{2}\times\sigma_{sky}\right)-AM\times E$.}}
\end{table}
The \eso\ observation of \zsfs\ in the $\textrm{R\_SPECIAL}$ band provides a deeper optical limit than previously achieved with \lbt\ \citep{2023A&A...674A.155K}. In Fig.~\ref{fig_vlt_img}, we mark the positions of nearby field sources detected using \texttt{SEXTRACTOR} \citep{1996A&AS..117..393B}, as well as the position of the target determined from \eros\ (blue circle) and \xmm\ (green circle; Table~\ref{tab_pos}). Clearly, there is no sign of a source in this blank region: the closest optical object is at a $>8\sigma$ separation from the well-determined EPIC pn position of the X-ray source.

The $5\sigma$ detection limit of the FORS2 image is 27.32\,mag (Table~\ref{tab_phot}). Using the relation presented in Sect.~3.4 of \cite{2024A&A...687A.251K} and the best-fit \texttt{2BBGABS} model flux of Model III in Table~\ref{tab_fitres}, we determined the X-ray-to-optical flux ratio limit to be above $5260$, an improvement of about one order of magnitude compared to the previous \lbt\ imaging.
\subsection{Gamma-ray analysis}
No counterpart is present within $\sim$1.1\degr\ of the position of the X-ray source in the fourth \fermi\ Source Catalogue Data Release 4 \citep[4FGL-DR4;][]{2022ApJS..260...53A,2023arXiv230712546B}. The closest entry is the blazar candidate \object{4FGL~J0658.2+2709}, with an energy flux of $1.8(4)\times10^{-12}$\,erg\,s$^{-1}$\,cm$^{-2}$ ($1\sigma$) in the $0.1-100$\,GeV energy range.

To search for gamma-ray emission from \zsfs, we retrieved Pass 8 \fermi-\textit{LAT} photon data accumulated between 2008 August 4 and 2024 June 11 within a region of interest (ROI) of 15\degr\ from the XINS position. The analysis was carried out with \texttt{fermitools}\footnote{\url{https://github.com/fermi-lat/Fermitools-conda}} following the guidelines from the \fermi\ Science Support Center. The photon files were filtered using \texttt{gtselect} within $0.1-500$\,GeV and zenith angles smaller than 90\degr. We only considered `source' events with \texttt{evclass=128} and \texttt{evtype=3}. Good-time intervals were applied with \texttt{gtmktime} and the filter expression \texttt{(DATA\_QUAL==1)\,\&\&\,(LAT\_CONFIG==1)}, ensuring that only data with the highest quality were used. Additionally, events registered during time intervals identified in the GTI extension of the 4FGL-DR4 catalogue as being associated with solar flares and gamma-ray bursts were removed from the analysis.

We carried out a binned likelihood analysis with \texttt{gtlike} to determine whether the XINS is detected as a point source \citep{2009ApJS..183...46A}. For this purpose, we adopted the recommended Galactic diffuse emission model, \texttt{gll\_iem\_v07.fits}, and isotropic spectral template, \texttt{iso\_P8R3\_SOURCE\_V3\_v1.txt}, and created a source model XML file with the locations, spatial and spectral shapes of the 4FGL sources within 25\degr\ from the XINS, which may contribute photons to the analysis at different energies. Count and source maps as well as the mission exposure and livetime were generated using \texttt{gtbin, gtsrcmaps, gtexpcube2} and \texttt{gtltcube}. 

The normalisation parameters of all bright and variable sources within 10\degr\ of the X-ray position, as well as the spectral parameters of the diffuse components, were left free during fitting; the parameters of the remaining 4FGL sources were kept fixed at their catalogued values. Altogether, the contribution of 277 \fermi\ sources, including four extended, was taken into account in the computation. 
Finally, we adopted a two-step minimisation procedure with \texttt{gtlike}, which is recommended for faint sources. Initially, we fitted the data with a tolerant convergence criterion using the MINUIT optimiser. This was followed by a more precise parameter estimation with NEWMINUIT over the initial results, using a strict fit convergence threshold.

Before introducing a putative source at the position of the XINS, we first performed fits to replicate the flux of sources catalogued in the 4FGL-DR4, which is based on a slightly shorter time baseline, and to achieve flat residuals across all energies. As an initial approximation, we then modelled the spectrum of the tentative gamma-ray counterpart of the INS using a simple power-law over the full sixteen-year time baseline of the analysis.

Consistent with the 4FGL-DR4 results, the likelihood analysis yields a low test statistic (TS) value of 6 at the position of the XINS, confirming the absence of a gamma-ray counterpart. Using the best-fit likelihood model, we derived upper limits for the pulsar's photon flux with the \texttt{UpperLimits} Python module in \texttt{fermitools}, allowing both the power-law normalisation and photon index to vary freely. The resulting 95\% confidence upper limit, $7 \times 10^{-9}$ photons s$^{-1}$ cm$^{-2}$, corresponds to an energy flux of $6.8\times10^{-13}$\,erg\,s$^{-1}$\,cm$^{-2}$ in the $0.1-100$\,GeV energy range.

Despite the lack of a spatial detection, we searched for pulsed gamma-ray emission using the X-ray timing solution from the \xmm\ and \nicer\ campaigns (see Section~\ref{sec_xraytiming}). We used an aperture radius of 5\degr\ for events above 100\,MeV. The photons were barycentred using the source coordinates listed in Table~\ref{tab_pos} and the JPL~DE405 solar system ephemeris. 
The short time baseline of the X-ray ephemeris ($\sim$126\,days) resulted in too few gamma-ray photons ($\sim$5,000) for a meaningful timing analysis. Therefore, we extended the search to cover the entire time baseline of the analysis. We adopted conservative $\pm3\sigma$ ranges in $\nu$ and $\dot{\nu}$ to account for all plausible values. To assess the significance of a potential signal, we used an implementation of the H-test optimised for detecting faint gamma-ray pulsars in \fermi-\textit{LAT} data\footnote{Python script \texttt{add\_weights.py} available at \url{https://fermi.gsfc.nasa.gov/ssc/data/analysis/user}}. This approach employs the `simple weights' approximation described by \citet{2019A&A...622A.108B,2019ApJ...871...78S}.
The searches yielded no significant modulation.
\section{Discussion\label{sec_disc}}
In this work, we report the results of a dedicated multi-wavelength follow-up campaign targeting the XINS candidate \zsfs. The deep \eso\ $R$-band observation confirms the compact nature of the source. The absence of counterparts brighter than 27.3\,mag ($5\sigma$) sets an extremely high X-ray-to-optical flux ratio ($>5260$), effectively ruling out more common soft X-ray emitters such as AGN, cataclysmic variables, or the active coronae of late-type stars. At the same time, this limit is still small compared to the X-ray-to-optical flux ratios of XINSs \citep[e.g.~the optical counterparts of the predominantly thermally emitting `magnificent seven' imply $\log(f_\mathrm{X}/f_\mathrm{opt}) \geq 4$; see Fig.~7 in][]{2024A&A...683A.164K}. Deeper optical observations of \zsfs\ will be necessary to uncover its optical counterpart.

At X-ray energies, the \xmm\ and \nicer\ observations have both unveiled the spin period of the neutron star at approximately 261\,ms (with a pulsed fraction of $(13.7 \pm 2.8)$\% in the EPIC pn data). The coherent timing analysis of the two epochs defines the pulsar spin-down rate at $\dot{P} = 6^{+11}_{-4} \times 10^{-15}$\,s\,s$^{-1}$. However, further observations and a longer time baseline are needed to unambiguously determine a timing solution. 
\begin{figure}[t]
\begin{center}
\includegraphics[width=\linewidth]{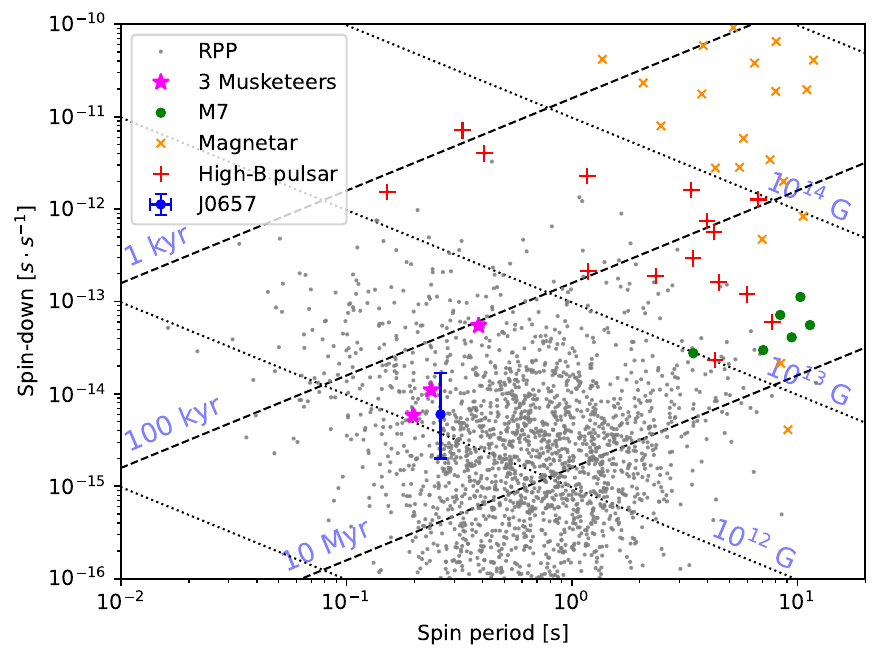}
\end{center}
\caption{Diagram of spin-down rate vs.~spin period of the pulsars from the ATNF catalogue \citep[version: 2.1.1;][]{2005AJ....129.1993M}. The location of \zsfs\ is indicated as a blue data point with error bars. We highlight the position of the `three musketeers' (\tmzsfs, \tmgem, and \tmozff; magenta), high magnetic field spin-powered pulsars \citep[red; same selection as in][]{2024A&A...683A.164K}, the `magnificent seven' (green), and magnetars (orange).} 
\label{fig_ppdot}
\end{figure}

In studying the phase-integrated X-ray spectrum of the source, we found that a \texttt{PL} continuum with a steep spectral slope of 4-5 or thermal models (two \texttt{BB} components at $kT_1 \sim 90$\,eV and $kT_2 \sim 220$\,eV, or an \texttt{NSA} at $\log(T/\mathrm{K}) \sim 5.8$ and a hot \texttt{BB} at $kT \sim 250$\,eV) fitted the continuum of \zsfs\ well. In all cases, an absorption feature around $\sim$0.3\,keV with a width of $\sim$50\,eV is necessary for a good fit. This feature was detected with high significance, with false-negative and false-positive rates of 6.2\% and 2.7\%, respectively. However, the line correlates strongly with the column density, allowing only upper limits for $\nh$ to be derived. Given the steep power-law slope, thermal models appear more physically plausible.

In Fig.~\ref{fig_ppdot} we show the source in the $(P,\dot{P})$ diagram of INSs from the ATNF pulsar catalogue \citep[version: 2.1.1; ][]{2005AJ....129.1993M}. Under the usual assumption of magnetic dipole braking in vacuum \citep{1969ApJ...157.1395O}, the magnetic field and characteristic age of spin-down, $B_\mathrm{dip} = 3.2\times10^{19}\sqrt{P\dot{P}}$\,G and $\tau_\mathrm{ch} = P(2\dot{P})^{-1}$\,s, are estimated as $1.2^{+0.9}_{-0.6} \times 10^{12}$\,G and $0.7^{+1.7}_{-0.5}$\,Myr, respectively, adopting the sample point distributions in Section~\ref{sec_xraytiming}. These values are typical of a middle-aged RPP.

\begin{figure}[t]
\begin{center}
\includegraphics[width=\linewidth]{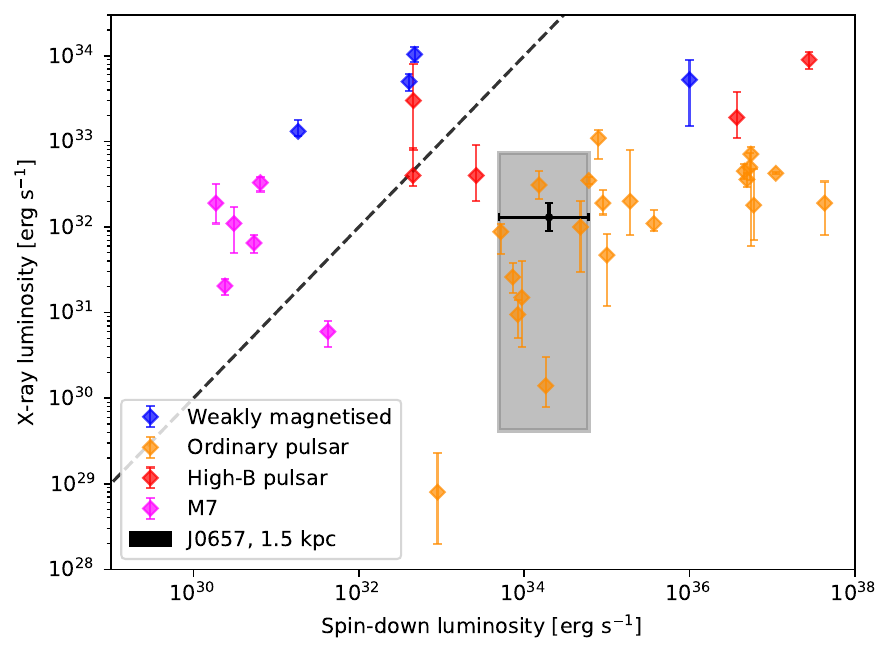}
\end{center}
\caption{X-ray vs.~spin-down luminosity diagram showing XINSs and pulsars with notable thermal emission \citep{2020MNRAS.496.5052P}. The black point with error bars represents the $1\sigma$ confidence region for \zsfs\ at a distance of 1\,kpc, based on \xmm\ and \nicer\ timing data (Sect.\ref{sec_xraytiming}). The shaded grey area shows the $1\sigma$ region for distances between 0.1\,kpc and 3\,kpc. The dashed line marks the identity line.
}
\label{fig_xlum}
\end{figure}

Comparing spin-down luminosity to X-ray luminosity can provide insights into the source's nature. This is because highly magnetised INSs can exhibit X-ray luminosities that exceed the energy available from spin-down, unlike spin-powered pulsars. Using the magnetic dipole braking model, we estimate the spin-down luminosity of \zsfs\ to be $\dot{E} = 6.3 \times 10^{46} (\dot{P}P^{-3})$ g\,cm$^2$ = $2.0^{+4}_{-1.5} \times 10^{34}$\,erg\,s$^{-1}$. The X-ray luminosity was calculated using the best-fit double blackbody model III in Table~\ref{tab_fitres} and the Stefan-Boltzmann law. To estimate the distance to the source, we applied the upper limit on the hydrogen column density ($\nh$) and used the 3D-$\nh$ Tool \citep{2024arXiv240303127D}, resulting in a distance upper limit of $< 1.5$\,kpc. In Fig.~\ref{fig_xlum}, we present the $1\sigma$ confidence regions and the positions of various thermally emitting INSs from \citet{2020MNRAS.496.5052P}. The diagram illustrates that, even at a distance of 1.5\,kpc, the absolute X-ray luminosity of \zsfs\ is roughly 100 times less than its spin-down luminosity, placing it comfortably within the region occupied by RPPs.

Young RPPs, such as the Crab pulsar, predominantly emit non-thermal X-rays that overshadow thermal radiation from their cooling surfaces \citep[see e.g.][for an overview]{2009ASSL..357...91B}. Older pulsars (ages greater than a few million years) show a weak thermal component, likely from small heated polar caps, alongside the dominant power-law emission. For RPPs aged around $10^5 - 10^6$ years, thermal emission is expected in soft X-rays (less than $1-2$\,keV), with magnetospheric emission dominating above 2\,keV. The spectrum of \zsfs\ suggests it may be an intermediately aged RPP, as it requires both cold and warm blackbody components to fit the data. This is consistent with models for the `three musketeers', where, additionally, power-law tails with spectral indices $\alpha = 1.7-2.1$ were detected at low flux levels ($0.3-1.7$\% of the source luminosity; \citealt{2005ApJ...623.1051D}). Similarly, the upper limit for \zsfs\ is $(2.1 \pm 1.8)$\%, indicating that any potential power-law tail is consistent with these faint levels.

Magnetospheric emission from RPPs may be detectable in both radio and gamma-ray wavelengths. Therefore, even if current X-ray observations are too shallow to identify non-thermal emission processes, detecting the source at other wavelengths might provide indirect evidence of such processes. Notably, the \fast\ observation of \zsfs\ failed to detect any modulation, resulting in a stringent $10\sigma$ upper limit of $1.4$\,$\mu$Jy at 1.5\,GHz. This result is reminiscent of the behaviour observed in \tmgem, which, despite numerous radio campaigns, has only shown sporadic claims of detection at low frequencies \citep[$<100$\,MHz; see][and references therein]{2015ARep...59..183M,2015ApJ...815..126M}. In contrast, \tmgem\ stands out as one of the most prominent gamma-ray sources, being among the first ever identified \citep[][]{1996ARA&A..34..331B}.

Given the similarities between the X-ray properties of \zsfs\ and \tmgem, we examined \fermi-\textit{LAT} data for a potential gamma-ray counterpart but found no significant spatial or pulsed signals over sixteen years of observations. We note that the X-ray timing solution from Section~\ref{sec_xraytiming} was extrapolated over the entire time baseline of the gamma-ray analysis, as there were few, if any, gamma-ray photons that could plausibly be attributed to \zsfs. At the most likely distance range of $0.7-1.5$\,kpc (Table~\ref{tab_fitres}), the 95\% c.l.~upper limit on the photon flux corresponds to a gamma-ray luminosity of $L_\gamma < (0.4-1.8) \times 10^{32}$ erg s$^{-1}$ in the $0.1-100$\,GeV energy range, assuming isotropic emission. This range lies at the lower end of the distribution for non-recycled pulsars with similar spin-down power \citep{2023ApJ...958..191S}. 

\begin{figure}[t]
\begin{center}
\includegraphics[width=\linewidth]{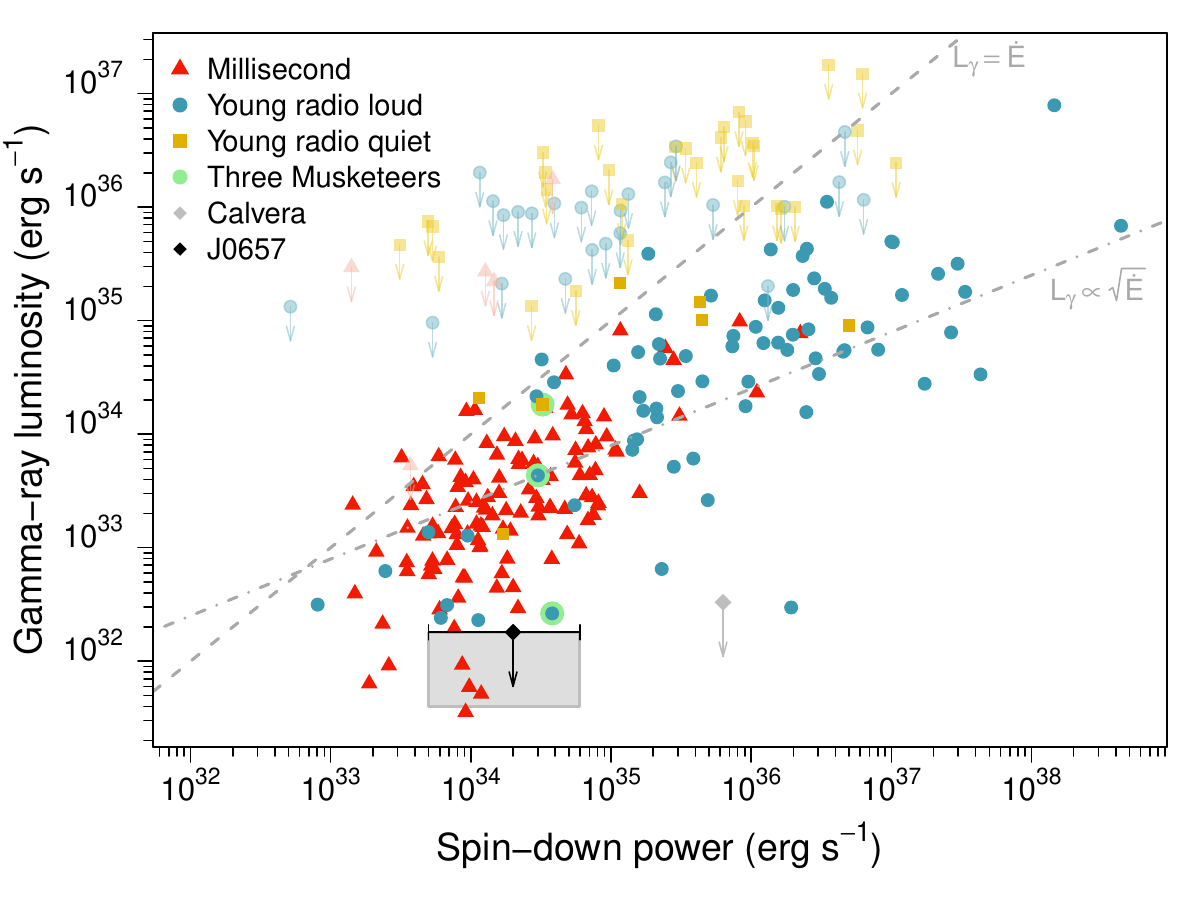}
\end{center}
\caption{Gamma-ray luminosity ($0.1-100$\,GeV) vs.~spin-down power of pulsars from the Third \fermi-\textit{LAT} catalogue \citep[adapted from][]{2023ApJ...958..191S}. The black diamond with error bars and the shaded grey area represent the 95\% c.l.~upper limit on the energy flux of the target, $6.8 \times 10^{-13}$ erg s$^{-1}$ cm$^{-2}$, at a distance range of $0.7-1.5$\,kpc. The positions of the `three musketeers' and \calvera\ are highlighted (see legend). The upper diagonal line shows where 100\% of spin-down power is converted to gamma-ray flux, while the lower diagonal line represents the heuristic relation $L_\gamma\propto\sqrt{\dot{E}}$, for reference.}
\label{fig_gammalum}
\end{figure}

The corresponding luminosity upper limit is illustrated in Fig.~\ref{fig_gammalum}, alongside the 274 catalogued pulsars, including millisecond ($P < 30$\,ms), radio-loud ($S_{1400} > 30$\,mJy, where $S_{1400}$ is the radio flux density at 1400\,MHz) and radio-quiet gamma-ray pulsars and their upper limits. The XINS's efficiency in converting spin-down power to gamma-ray luminosity, $\eta \equiv L_{\gamma}/\dot{E} < 0.9\%$, is also notably lower than most of non-recycled pulsars with similar $\dot{E}$. The position of the high-latitude X-ray pulsar \calvera\ \citep[\calveraf; e.g.][and references therein]{2013ApJ...778..120H,2021ApJ...922..253M} is also indicated. Estimated to be at a distance of $3.3$\,kpc, \calvera\ is a spin-powered pulsar that may have been born at a significant height above the Galactic disk. Despite its relatively high $\dot{E}$, it does not exhibit radio or gamma-ray emissions. Given these characteristics, it is speculated that \calvera\ could be a descendant of the central compact object class -- a group of enigmatic, radio-quiet young neutron stars found in supernova remnants \citep[see e.g.][]{2017JPhCS.932a2006D}.

A potential explanation for the absence of magnetospheric emission in certain INSs is the misalignment of the beamed non-thermal emission with the observer's line of sight. This idea has been proposed to account for the lack of detectable radio emission in nearby, long-spin-period XINSs \citep{2009ApJ...702..692K}. Nearly half of non-recycled gamma-ray pulsars are radio-faint (70 out of 150), and 95\% of radio pulsars with similar $\dot{E}$ remain unseen at GeV energies \citep[see][for details]{2023ApJ...958..191S}. While the narrow radio beam is typically aligned with the INS polar regions, the difference between radio and gamma-ray emissions suggests that gamma-rays originate from a different region in the outer magnetosphere. Models \citep[e.g.][]{1998ApJ...498..327C, 2006MNRAS.366.1310T, 2008MNRAS.386..748T} propose that gamma-rays cover a wider angle than the radio beam, which could explain the phenomenon of radio-quiet pulsars among RPPs.

In addition, both beams may miss the observer. The population synthesis simulations of \citet{2020MNRAS.497.1957J}, although applied to the young and energetic portion of the spin-powered pulsar population with $\dot{E} > 10^{35}$\,erg\,s$^{-1}$, suggest that such a population of `invisible' pulsars could account for 25\% of the total. Alternatively, the absence of non-thermal emission may indicate unusual properties in the neutron star's magnetosphere, rather than merely an unfavourable viewing geometry. \citet{2011ApJ...738..114R} proposed a class of sub-luminous gamma-ray pulsars with $\dot{E} > 10^{34}$\,erg\,s$^{-1}$ and $d < 2$\,kpc, characterised by low gamma-ray efficiency, possibly due to a lower-altitude emission component. The XINS \zsfs, the `musketeer' \tmzsfs, and \calvera\ all belong to this category.

Absorption features in various INSs are typically linked to cyclotron resonances of charged particles in the neutron star's magnetic field \citep{2019A&A...622A..61S}, bound-bound or bound-free transitions in the atmosphere \citep[e.g.][]{2007Ap&SS.308..191V}, or inaccuracies in temperature modelling \citep{2014MNRAS.443...31V}. If cyclotron resonances are the cause, the magnetic field strength for the first harmonic is given by  $B_\mathrm{cyc} = (z+1)\frac{mE}{\hbar q}$, where $m$ is the particle mass, $E$ is the energy, and $q$ is the charge. Using this formula with the absorption feature's best-fit energy, we find $B_\mathrm{cyc,p^+} = 6.5^{+1.1}_{-0.6} \times 10^{13}$\,G for protons and $B_\mathrm{cyc,e^-} = 3.56^{+0.6}_{-0.28} \times 10^{10}$\,G for electrons. These values differ from the typical $10^{12}$\,G seen in intermediately aged RPPs, as indicated by spin-parameter limits. The higher field strength for proton cyclotron resonances indicates a more magnetised region, likely near the INS surface. In contrast, the lower field strength for electron cyclotron resonances suggests a region farther from the surface, where the magnetic field is weaker, as has been proposed for \tmzsfs\ \citep{2018ApJ...869...97A}.

Alternatively, the absorption feature might be due to bound-bound or bound-free transitions in the INS atmosphere or surface. If we assume that the line is caused by hydrogen ionisation, the derived magnetic field strength is $3.1^{+2.8}_{-0.9} \times 10^{13}$\,G, which exceeds both timing constraints and typical values for RPPs. However, transitions in elements with higher atomic numbers could explain the observed line, as these can occur at higher energies than hydrogen \citep{2007MNRAS.377..905M, 2008MNRAS.383..161M}. We also note that deeper X-ray observations might reveal additional features if transitions in higher \(Z\) elements are present.

Finally, the absorption feature might result from inadequate modelling of the surface temperature distribution. \cite{2014MNRAS.443...31V} suggested that this could explain similar features in some INSs, typically seen between two adjacent blackbody components with temperatures differing at least by a factor of 1.5 to 2. While this applies to the \texttt{BB} components in fit III (Table~\ref{tab_fitres}), the observed line energy is too low to fall between these components (Wien's law suggests the BB peaks at $\sim$0.5\,keV and $\sim$1.1\,keV, respectively). This discrepancy might indicate the presence of additional emission components at very soft X-rays or far-ultraviolet, which would require further observations.

Additional insights into the nature of the line could be gained from phase-dependent studies. Cyclotron resonances, for instance, may show phase-dependence, as suggested for a feature in \tmzsfs\ \citep{2018ApJ...869...97A,2022A&A...661A..41S}. However, the EPIC pn observation only permits splitting the rotation into two phase bins, and the contamination in the \nicer\ data is too strong for a reliable phase-resolved analysis. Statistically, phase-resolved spectra fit well without an absorption component (see Table~\ref{tab_fitphres}), and including the component yields similar line parameters to those in the phase-averaged case. Due to the limited photon count, uncertainties in the line parameters are too large to detect phase-evolution. Therefore, higher signal-to-noise X-ray data with better timing resolution are needed to investigate \zsfs's phase-resolved properties.
\section{Summary and conclusions\label{sec_concl}}
The emission properties of \jzsfs\ (\jzsfsatnf) are consistent with those of a middle-aged spin-powered pulsar. However, it is rare within the observed population to find pulsars of this type without detectable pulsed emission in both radio and gamma-ray energies. While this absence is most likely due to an unfavourable viewing geometry, the source's rarity makes it particularly valuable for studying magnetospheric processes and understanding potential deviations in emission geometry. Observations at low radio frequencies could help determine whether mechanisms similar to those in \tmgem\ and other radio-quiet gamma-ray pulsars are at play. Further X-ray observations could also provide key insights: by enabling searches for faint non-thermal emission, refining the timing solution to constrain the braking rate (critical for gamma-ray pulsation searches), and facilitating a more reliable phase-resolved spectral analysis to probe the origin of the absorption feature. Ultimately, the discovery and detailed study of this XINS highlight the unique potential of X-ray observations to uncover rare sources that conventional pulsar surveys often miss.
\begin{acknowledgements}
The authors would like to express their thanks to Wu Xiang Ping and Peng Jiang for the prompt scheduling of the \fast\ DDT observation. We thank Sergei~Popov for valuable discussions regarding the nature of our target during the 2024 XMM-Newton Workshop. We also thank the referee, David Smith, for his suggestions for improvement, and the \fermi\ Helpdesk, particularly Nestor Mirabal, for their extensive guidance while analysing LAT observations.

This work was funded by the Deutsche Forschungsgemeinschaft (DFG, German Research Foundation) -- 414059771.
AMP acknowledges the Innovation and Development Fund of Science and Technology of the Institute of Geochemistry, Chinese Academy of Sciences, the National Key Research and Development Program of China (Grant No. 2022YFF0503100), the Strategic Priority Research Program of the Chinese Academy of Sciences (Grant No. XDB 41000000), and the Key Research Program of the Chinese Academy of Sciences (Grant NO. KGFZD-145-23-15).

IT gratefully acknowledges the support by Deutsches Zentrum für Luft- und Raumfahrt (DLR) through grant 50\,OX\,2301.

This research has made use of data and/or software provided by the High Energy Astrophysics Science Archive Research Center (HEASARC), which is a service of the Astrophysics Science Division at NASA/GSFC.

This work has used the data from the Five-hundred-meter Aperture Spherical radio Telescope (FAST). FAST is a Chinese national mega-science facility, operated by the National Astronomical Observatories of Chinese Academy of Sciences (NAOC).

Based on observations made with ESO Telescopes at the La Silla Paranal Observatory under programme ID 111.259R.001

This work is based on data from eROSITA, the soft X-ray instrument aboard SRG, a joint Russian-German science mission supported by the Russian Space Agency (Roskosmos), in the interests of the Russian Academy of Sciences represented by its Space Research Institute (IKI), and the Deutsches Zentrum für Luft- und Raumfahrt (DLR). The SRG spacecraft was built by Lavochkin Association (NPOL) and its subcontractors, and is operated by NPOL with support from the Max Planck Institute for Extraterrestrial Physics (MPE).

The development and construction of the eROSITA X-ray instrument was led by MPE, with contributions from the Dr. Karl Remeis Observatory Bamberg \& ECAP (FAU Erlangen-Nuernberg), the University of Hamburg Observatory, the Leibniz Institute for Astrophysics Potsdam (AIP), and the Institute for Astronomy and Astrophysics of the University of Tübingen, with the support of DLR and the Max Planck Society. The Argelander Institute for Astronomy of the University of Bonn and the Ludwig Maximilians Universität Munich also participated in the science preparation for eROSITA.

This work made use of Astropy:\footnote{http://www.astropy.org} a community-developed core Python package and an ecosystem of tools and resources for astronomy \citep{astropy:2013, astropy:2018, astropy:2022}. This research made use of ccdproc, an Astropy package for image reduction \citep{matt_craig_2017_1069648}.

We derive posterior probability distributions and the Bayesian evidence with the nested sampling Monte Carlo algorithm MLFriends \citep{2014A&A...564A.125B,2019PASP..131j8005B} using the UltraNest\footnote{\url{https://johannesbuchner.github.io/UltraNest/}} package \citep{2021JOSS....6.3001B}.
\end{acknowledgements}
\bibliographystyle{aa}
\bibliography{ref_insc_j0657}
\end{document}